\documentclass{moriond}

\usepackage{amsmath}
\usepackage{graphicx}
\usepackage{subcaption}

\def\be{\begin{equation}}
\def\ee{\end{equation}}
\def\bea{\begin{eqnarray}}
\def\eea{\end{eqnarray}}

\def\panscales{\textsc{PanScales}}

\newcommand{\Photo}{\includegraphics[height=35mm]{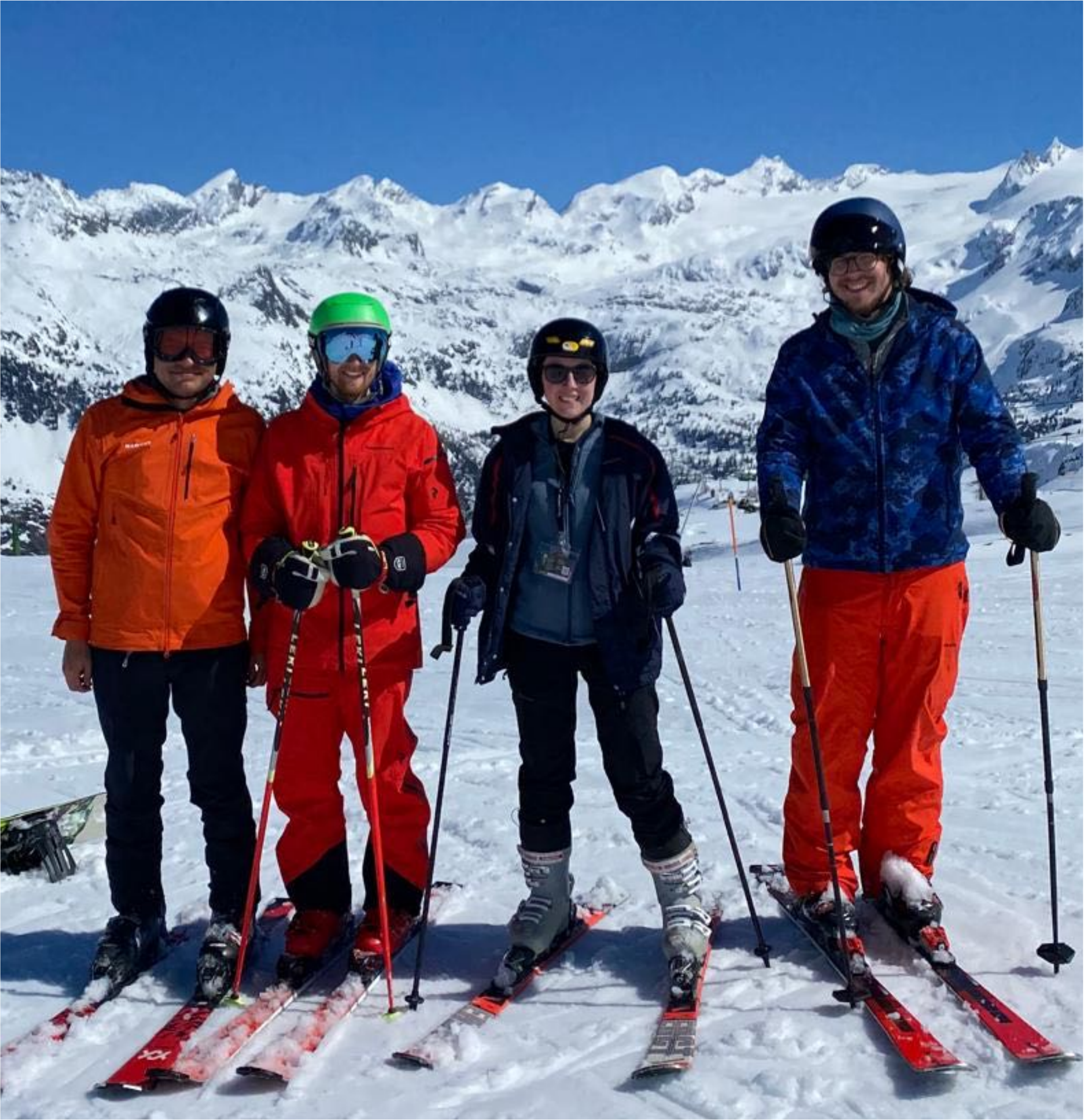}}

\begin{document}
\vspace*{4cm}
\title{
Building Next-to-Next Leading Logarithmic parton showers:
the PanScales recipe
 }

\author{Silvia Ferrario Ravasio}

\address{Theoretical Physics Department, CERN, CH-1211 Geneva 23, Switzerland}

\maketitle\abstracts{
  Parton showers lie at the core of Shower Monte Carlo event generators, the default theoretical tools used to interpret collider data.
  In these proceedings, we summarise the strategy of the \panscales{} collaboration that led to the attainment of the first demonstrably next-to-next-to-leading-logarithmic accurate parton showers.\cite{vanBeekveld:2024wws}
}

\section{Introduction}
Collider phenomenology requires the description of physical processes across a broad range of energy scales, from the hard momentum transferred during the collision \( Q \sim 100\,\mathrm{GeV} \)--\(1\,\mathrm{TeV} \), down to soft scales \( \Lambda \sim 1\,\mathrm{GeV} \), where free partons confine into hadrons.  
The presence of such disparate scales leads to logarithmically enhanced contributions at all orders in the strong coupling constant \( \alpha_s \).
These contributions can be predicted either through analytically resummed calculations or via parton-shower algorithms.
While analytic resummation techniques can achieve remarkable accuracy
(see e.g.\ Refs.~\cite{Duhr:2022yyp,Chen:2022cgv,Camarda:2023dqn}), their applicability is limited to relatively simple observables.
Moreover, the systematic inclusion of non-perturbative corrections within such frameworks remains an open challenge.
By contrast, parton-shower simulations offer a fully exclusive event description and are implemented within Shower Monte Carlo (SMC) generators,\cite{Bahr:2008pv,Bierlich:2022pfr,Sherpa:2024mfk} which also incorporate sophisticated phenomenological models for non-perturbative effects.
Thus, although analytically resummed predictions offer higher formal accuracy, it is SMC simulations that are ultimately compared with experimental data.

Parton showers routinely used to interpret LHC data typically account only for the dominant logarithmically enhanced corrections~\cite{Dasgupta:2018nvj}, known as leading logarithms (LL), i.e.\ terms of the form \(L(\alpha_s L)^n\), where \(L = \ln(Q/\Lambda)\).
%
Next-to-leading logarithms (NLL, \((\alpha_s L)^n\)) often give rise to 30-50\% corrections.
Recent years witnessed the development of many NLL-accurate parton showers.\cite{Dasgupta:2020fwr,Nagy:2020rmk,Forshaw:2020wrq,Herren:2022jej,vanBeekveld:2022zhl,vanBeekveld:2023chs,Preuss:2024vyu}
Their integration into fully fledged SMC simulations remains an ongoing endeavour.
A particularly notable milestone is the attainment of NNLL accuracy for broad classes of observables in \(e^+e^-\) collisions.\cite{FerrarioRavasio:2023kyg,vanBeekveld:2024wws}
In these proceedings, we summarise the \panscales{} collaboration's strategy for constructing parton showers with accuracy beyond LL, based on ingredients systematically derived from analytic resummation.\cite{Catani:1990rr,Banfi:2018mcq,Dasgupta:2021hbh,vanBeekveld:2023lsa}

\section{Building an NLL Parton Shower}
Dipole showers~\cite{Gustafson:1987rq} are the most commonly used showers, due to their advantages in matching with fixed-order calculations.
%
In the large-number-of-colours limit, a dipole consists of a pair of colour-connected partons, and an emission corresponds to its splitting into two dipoles.
The splitting procedure is repeated independently for each dipole until no further emissions are possible above a non-perturbative cut-off  $\Lambda$.
Dipole showers rely on three key components:
(1) an effective matrix element for the branching of a two-parton system into three;
(2) a momentum mapping to reshuffle the momenta when a new particle is created;
(3) an ordering variable, typically transverse momentum or virtuality.

\begin{figure}[t]
  \begin{subfigure}[t]{0.42\textwidth}
    \centering
    \raisebox{0.5cm}{
    \includegraphics[width=\textwidth]{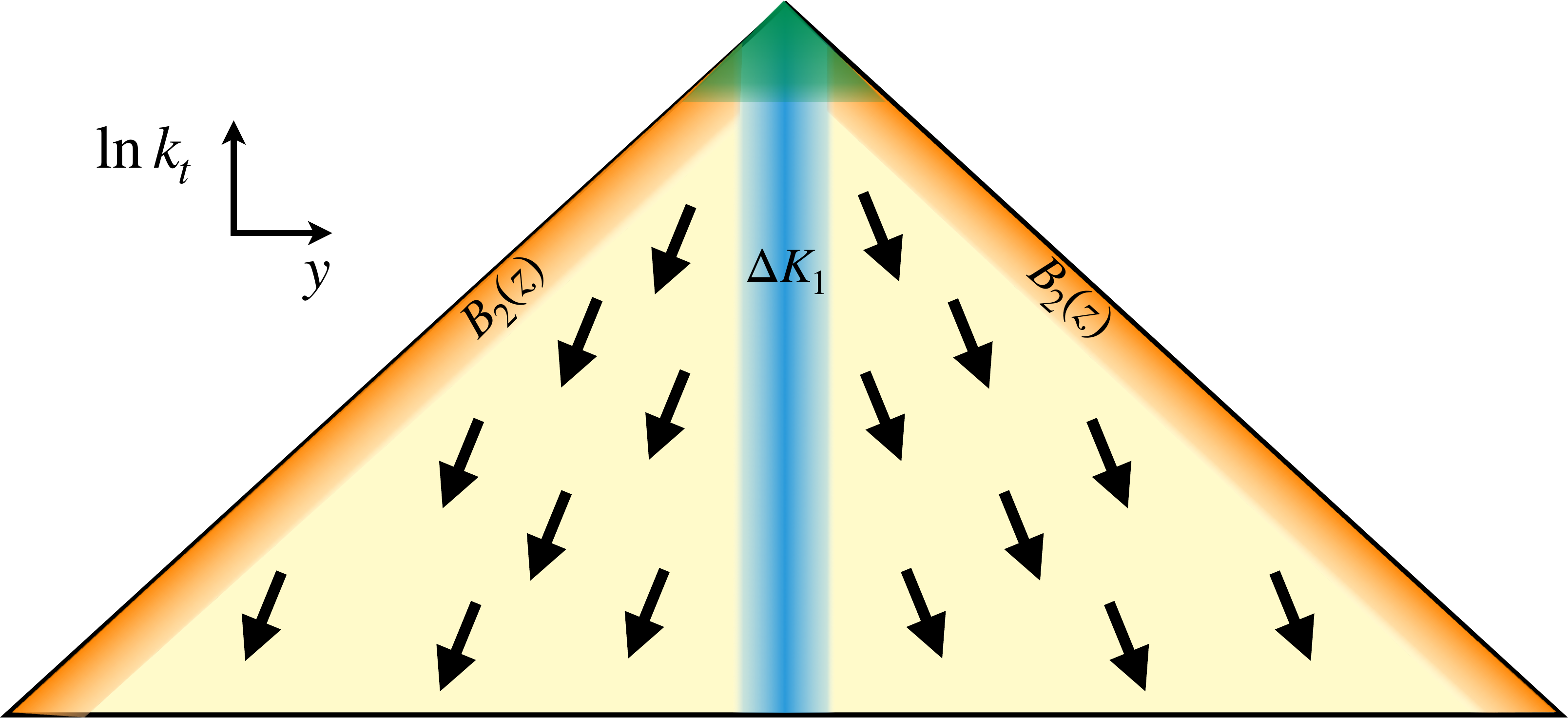}}
  \caption{
  Lund-plane representation of the phase space of an emission in terms of its transverse momentum $k_t$ and rapidity $y$. 
   }
  \label{fig:LundPlane}
  \end{subfigure}
  \hfill
  \begin{subfigure}[t]{0.54\textwidth}
    \centering
    \includegraphics[width=\textwidth]{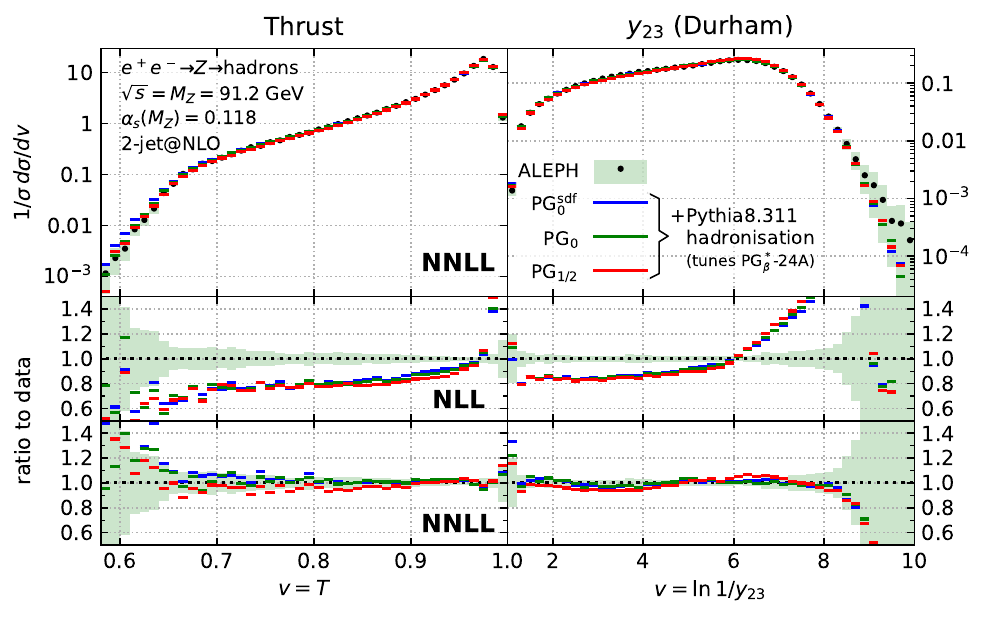}
    \caption{
    Selected $e^+e^-$ event shapes compared to ALEPH data.\cite{ALEPH:2003obs} The lower (middle) panel shows the ratios of \panscales{} NNLL (NLL) shower variants to data.
    }
    \label{fig:evtsh}
  \end{subfigure}
\end{figure}

How these elements define the logarithmic accuracy of a parton shower can be understood by introducing the Lund plane,\cite{Andersson:1988gp} which describes the radiation phase space in terms of transverse momentum $k_t$ and rapidity $y$, as shown in Fig.~\ref{fig:LundPlane}.
For NLL accuracy, the parton shower matrix element must reproduce the correct tree-level QCD matrix element in the limit where the emission is collinear and/or soft.  
Higher-order effects stemming from unresolved emissions are incorporated using the Monte Carlo scheme for \(\alpha_s\):  
\begin{equation}
\alpha_s^{\rm MC} = \alpha_s(k_t)\left[ 1 + \frac{\alpha_s(k_t)}{2\pi}K_1 \right],\quad \mbox{with }K_1 = \left( \frac{67}{18} - \frac{\pi^2}{6} \right) C_A - \frac{10}{9} T_R n_f,
\label{eq:KMC}
\end{equation}
where \(\alpha_s(k_t)\) is the \(\overline{\rm MS}\) coupling constant, evaluated at the transverse momentum of the emission, and \(K_1\) is a universal constant.\cite{Catani:1990rr}  
The running allows the resummation of all \((\alpha_s L)^n\) contributions for emissions that are either hard-collinear (orange region in Fig.~\ref{fig:LundPlane}) or soft large-angle (blue region).  
For soft-collinear radiation (yellow area of Fig.~\ref{fig:LundPlane}), it is necessary to include the constant \(K_1\), as well as to promote the running of \(\alpha_s\) to two loops.

At NLL accuracy, one must also account for contributions from multiple soft-collinear emissions that are widely separated in angle but have comparable hardness.  
Colour coherence implies that the matrix element factorises into a product of independent  emissions.  
Thus, the interplay between the shower ordering variable and its kinematic map must preserve the coherence of multiple independent soft-collinear emissions.  
This is precisely where conventional dipole showers exhibit limitations.\cite{Dasgupta:2018nvj} 
Addressing this issue is central to the development of a new class of parton showers,\cite{Dasgupta:2020fwr,Nagy:2020rmk,Forshaw:2020wrq,Herren:2022jej,vanBeekveld:2022zhl,vanBeekveld:2023chs,Preuss:2024vyu} which are designed to achieve NLL accuracy.

\section{Going beyond NLL accuracy}

\paragraph{NNLL corrections from the hard region}
Matching parton showers with NLO calculations~\cite{Frixione:2002ik,Nason:2004rx} has been employed for more than 20 years to improve the description of the non-logarithmically enhanced region of phase space, i.e.\ the top of the Lund plane.  
Ref.~\cite{Hamilton:2023dwb} presents the first systematic study of the interplay between NLO corrections and the logarithmic accuracy of parton showers. It demonstrates how NLO matching can be used to include the \(\alpha_s\) constant terms that enter the tower of NNLL contributions of the form \(\alpha_s(\alpha_s L)^n\).  
It is, however, essential that the phase space associated with the hardest emission--described using the exact tree-level matrix element--is not double-counted by subsequent emissions.  
The work in Ref.~\cite{Hamilton:2023dwb} was recently extended to include processes with incoming partons in Ref.~\cite{vanBeekveld:2025lpz}, which also introduces new, efficient NLO matching paradigms.

\paragraph{NNLL corrections from the soft region}
To achieve NNLL accuracy, it is necessary to include both real and virtual corrections to soft emissions.  
The implementation of real corrections involves the description of a pair of soft partons.  
In practice, for a parton shower that is NLL accurate and covers the full phase space, corrections to the double-soft emission rate are only required when the emitted partons are close in both transverse momentum and rapidity.  
To incorporate this correction, each time a parton \(i\) is emitted close in phase space to a previously emitted soft parton \(j\), the single-emission probability of \(i\) is multiplied by the ratio of the exact double-soft matrix element for producing \(i\) and \(j\) to the effective parton shower one.\cite{FerrarioRavasio:2023kyg}  

The inclusive NLO rate for a soft emission is given by \(1 + \alpha_s/(2\pi) K_1\), where \(K_1\) is the constant appearing in the Monte Carlo scheme for the coupling~\eqref{eq:KMC}.  
This constant originates from summing virtual corrections with the corresponding real corrections integrated over the gluon decay phase space while keeping the transverse momentum and rapidity of the parent gluon fixed.  
The choice of which kinematic variables are held fixed when integrating the real contribution defines the resummation scheme.  
In a generic NLL shower, the rapidity and transverse momentum of an emission can drift due to subsequent branchings, as illustrated by the arrows in Fig.~\ref{fig:LundPlane}.  
This leads to variations in the emission density across the central, large-angle region of the Lund plane.  
To account for this effect, we introduce a correction factor \(\Delta K_1(y)\), corresponding to the difference between the integral of the double-soft matrix element over the phase space defined by the parton shower mapping and that defined by the resummation scheme.\cite{FerrarioRavasio:2023kyg}  

The inclusion of \(\Delta K_1(y)\), together with the double-soft acceptance probability and NLO matching, is sufficient to capture all terms of the form \(\alpha_s(\alpha_s L)^n\) for observables not sensitive to collinear radiation, such as certain non-global observables.  
For these observables, the accuracy of the parton shower is identical to that of state-of-the-art analytic calculations.\cite{Banfi:2021xzn,Becher:2023vrh}

\paragraph{NNLL corrections from the soft-collinear region}
The results described in the previous section also apply to the soft-collinear region, where \(\Delta K_1 \to 0\), due to the fact that the shower is already NLL.   
For a soft-collinear emission, Eq.~\eqref{eq:KMC} must be extended to include both the three-loop running of \(\alpha_s\) and the NNLO inclusive correction \(K_2\), as computed, for example, in Ref.~\cite{Banfi:2018mcq}.  
At NLO, recoil effects due to multiple emissions do not affect the Lund-plane density in the bulk, because the rapidity and transverse momentum drifts \(\Delta y_{\rm sc}\) and \(\Delta \ln k_{t,{\rm sc}}\) of a soft-collinear emission are constant.  
However, the Lund-plane density is computed using an incorrect argument for \(\alpha_s\), which induces an incorrect \(\alpha_s^3\) term.  
As a result, the effective \(K_2\) used in the parton shower must include a correction term proportional to \(\Delta \ln k_{t,{\rm sc}}\) and the one-loop QCD beta function, to compensate for the transverse-momentum drift.\cite{vanBeekveld:2024wws}

\paragraph{NNLL corrections to global event shapes from the hard-collinear region}
We now describe how to improve the modelling of the hard-collinear region in order to achieve NNLL accuracy.  
Focusing on global observables, it is sufficient to include the average NLO correction to the emission of a hard-collinear parton, denoted \(B_2\).  
The inclusive \(\alpha_s\) correction \(B_2(z)\), expressed as a function of \(z = k_t e^{-y}\), can be obtained from Refs.~\cite{Dasgupta:2021hbh,vanBeekveld:2023lsa}. We compute its average by integrating over \(z\).  
Analogously to \(K_1\), the result derived in the resummation scheme must be corrected by a factor proportional to \(\Delta \ln z_{\rm sc} = \Delta \ln k_{t,{\rm sc}} + \Delta y_{\rm sc}\), to compensate for soft-collinear emissions migrating into the hard-collinear region.

\section{Concluding remarks}
The combination of the previously described ingredients enables the formulation of the first parton shower achieving NNLL accuracy for any global event shape, as well as for non-global observables primarily sensitive to soft emissions.
In Fig.~\ref{fig:evtsh}, we show that including these corrections significantly improves agreement with data in \( e^+e^- \) collisions for selected event shapes.

The next goals of the \panscales{} collaboration are the attainment of the correct triple-collinear dynamics, necessary for general NNLL accuracy, as well as NNLL accuracy for processes with incoming partons and multiple hard jets in the final state.

\section*{Acknowledgements}
SFR thanks G.~P.~Salam for comments on the manuscript.
The results presented in these proceedings have been obtained in collaboration with M.~van Beekveld, M.~Dasgupta, F. Dreyer, B.~K.~El-Menoufi, K.~Hamilton, J.~Helliwell, A.~Karlberg, R. Medves, P.~F.~Monni, G.~P.~Salam, L.~Scyboz, A. Soto Ontoso, G. Soyez, R. Verheyen and S. Zanoli.

\end{document}